\documentclass[aip, reprint]{revtex4-1}

\usepackage{amsmath}
\usepackage{amssymb}
\usepackage[english]{babel}
\usepackage{dcolumn}
\usepackage[T1]{fontenc}
\usepackage{graphicx}
\usepackage[utf8]{inputenc}

\draft 

\begin{document}


\title[Production of highly charged ions of rare species by laser-induced desorption inside an EBIT]{Production of highly charged ions of rare species by laser-induced desorption inside an electron beam ion trap} 



\author{Ch. Schweiger}
\email[]{christoph.schweiger@mpi-hd.mpg.de}
\affiliation{Max-Planck-Institut für Kernphysik, Saupfercheckweg 1, 69117 Heidelberg, Germany}

\author{C.M. König}
\affiliation{Max-Planck-Institut für Kernphysik, Saupfercheckweg 1, 69117 Heidelberg, Germany}
\affiliation{Fakultät für Physik und Astronomie, Universität Heidelberg, Im Neuenheimer Feld 226, 69120 Heidelberg, Germany}

\author{J.R. Crespo López-Urrutia}
\affiliation{Max-Planck-Institut für Kernphysik, Saupfercheckweg 1, 69117 Heidelberg, Germany}

\author{M. Door}
\affiliation{Max-Planck-Institut für Kernphysik, Saupfercheckweg 1, 69117 Heidelberg, Germany}

\author{H. Dorrer}
\affiliation{Institut für Kernchemie, Johannes Gutenberg-Universität Mainz, Fritz-Strassmann-Weg 2, 55128 Mainz, Germany}

\author{Ch. E. Düllmann}
\affiliation{Institut für Kernchemie, Johannes Gutenberg-Universität Mainz, Fritz-Strassmann-Weg 2, 55128 Mainz, Germany}
\affiliation{Helmholtz-Institut Mainz, Staudingerweg 18, 55128 Mainz, Germany}
\affiliation{GSI Helmholtzzentrum für Schwerionenforschung GmbH, Planckstraße 1, 64291 Darmstadt, Germany}

\author{S. Eliseev}
\affiliation{Max-Planck-Institut für Kernphysik, Saupfercheckweg 1, 69117 Heidelberg, Germany}

\author{P. Filianin}
\affiliation{Max-Planck-Institut für Kernphysik, Saupfercheckweg 1, 69117 Heidelberg, Germany}

\author{W. Huang}
\affiliation{Max-Planck-Institut für Kernphysik, Saupfercheckweg 1, 69117 Heidelberg, Germany}

\author{K. Kromer}
\affiliation{Max-Planck-Institut für Kernphysik, Saupfercheckweg 1, 69117 Heidelberg, Germany}
\affiliation{Fakultät für Physik und Astronomie, Universität Heidelberg, Im Neuenheimer Feld 226, 69120 Heidelberg, Germany}

\author{P. Micke}
\affiliation{Max-Planck-Institut für Kernphysik, Saupfercheckweg 1, 69117 Heidelberg, Germany}
\affiliation{Physikalisch-Technische Bundesanstalt, Bundesallee 100, 38116 Braunschweig, Germany}

\author{M. Müller}
\affiliation{Max-Planck-Institut für Kernphysik, Saupfercheckweg 1, 69117 Heidelberg, Germany}
\affiliation{Fakultät für Physik und Astronomie, Universität Heidelberg, Im Neuenheimer Feld 226, 69120 Heidelberg, Germany}

\author{D. Renisch}
\affiliation{Institut für Kernchemie, Johannes Gutenberg-Universität Mainz, Fritz-Strassmann-Weg 2, 55128 Mainz, Germany}
\affiliation{Helmholtz-Institut Mainz, Staudingerweg 18, 55128 Mainz, Germany}

\author{A. Rischka}
\affiliation{Max-Planck-Institut für Kernphysik, Saupfercheckweg 1, 69117 Heidelberg, Germany}

\author{R.X. Schüssler}
\affiliation{Max-Planck-Institut für Kernphysik, Saupfercheckweg 1, 69117 Heidelberg, Germany}

\author{K. Blaum}
\affiliation{Max-Planck-Institut für Kernphysik, Saupfercheckweg 1, 69117 Heidelberg, Germany}

\date{\today}

\begin{abstract}
This paper reports on the development and testing of a novel, highly efficient technique for the injection of very rare species into electron beam ion traps (EBITs) for the production of highly charged ions (HCI).
It relies on in-trap laser-induced desorption of atoms from a sample brought very close to the electron beam resulting in a very high capture efficiency in the EBIT.
We have demonstrated a steady production of HCI of the stable isotope $^{165}\mathrm{Ho}$ from samples of only $10^{12}$ atoms ($\sim$ 300~pg) in charge states up to 45+.
HCI of these species can be subsequently extracted for use in other experiments or stored in the trapping volume of the EBIT for spectroscopic measurements. 
The high efficiency of this technique expands the range of rare isotope HCIs available for high-precision nuclear mass and spectroscopic measurements.
A first application of this technique is the production of HCI of the synthetic radioisotope $^{163}\mathrm{Ho}$ for a high-precision measurement of the $Q_{\mathrm{EC}}$-value of the electron capture in $^{163}\mathrm{Ho}$ within the ``\textbf{E}lectron \textbf{C}apture in \textbf{Ho}lmium'' experiment~\cite{Gastaldo14, Gastaldo17} (\textsc{ECHo} collaboration) ultimately leading to a measurement of the electron neutrino mass with an uncertainty on the sub-eV level.

\end{abstract}


\maketitle 



\section{Introduction}
\label{sec:introduction}

Many experiments require access to highly charged ions (HCI) of species that cannot be found in nature and hence have to be synthesized in nuclear reactions at rare-ion-beam facilities~\cite{Kugler00, Gastaldo13} and research reactors~\cite{Dorrer18}.
These species can be produced only in very small quantities (rare species), often only in the sub nanogram region.
Applications requiring the production of HCI of these species include a direct test of the theory of special relativity~\cite{Rainville05, Jentschel18}, involving a precise measurement of the neutron binding energy in $^{36}\mathrm{Cl}$ by measuring the mass ratio of $^{35}\mathrm{Cl}$ and $^{36}\mathrm{Cl}$, as well as high-precision mass measurements of transuranium elements to establish new anchor points in $\alpha$ decay chains, thereby reducing the uncertainty in the masses of superheavy nuclides allowing the identification of nuclear shell closures~\cite{Eibach14}. 
High-precision mass measurements of HCI can furthermore support g-factor measurements~\cite{Kohler16} and searches for Dark Matter in high-resolution isotope shift measurements~\cite{Flambaum18, Antypas19, Manovitz19} using e.g. enriched samples of rare Ca isotopes.

One of these experiments is the ``\textbf{E}lectron \textbf{C}apture in \textbf{Ho}lmium'' experiment~\cite{Gastaldo14, Gastaldo17} (\textsc{ECHo} collaboration) aiming at a measurement of the neutrino mass with an uncertainty on the sub-eV level.
Since the discovery of neutrino oscillations~\cite{Fukuda98, Ahmad02}, establishing that neutrinos have a finite mass, the measurement of the absolute scale of the neutrino mass remains a challenging task due to the exclusively weak interaction with other standard model particles and the small absolute mass scale.
For an improvement towards the sub-eV level the ECHo collaboration calorimetrically measures a high statistics spectrum of the electron capture in $^{163}\mathrm{Ho}$.
In the analysis of this spectrum, the precise knowledge of the energy available for the decay, $Q_{\mathrm{EC}}$, from an independent source is required to investigate systematic effects in the calorimetric measurement.
As the $Q_{\mathrm{EC}}$ value corresponds to the mass difference between mother and daughter nuclides it can be directly accessed using a mass spectrometer.
The required uncertainty of 1~eV, corresponding to a relative uncertainty of $\frac{\delta Q_{\mathrm{EC}}}{m} \sim 6\!\cdot\!10^{-12}$, can currently only be reached using high-precision Penning-trap mass spectrometry~\cite{Blaum06, Dilling17}.
The high-precision Penning-trap mass spectrometer \textsc{Pentatrap}~\cite{Repp12} has recently shown its capability to reach the required uncertainty using HCI of xenon, rhenium and osmium\cite{Rischka19, Schussler19}.
$^{163}\mathrm{Ho}$ is a synthetic radioisotope with a half-life of 4570(25) years of which only small amounts can be produced by neutron irradiation of an enriched $^{162}\mathrm{Er}$ target in a research reactor and subsequent chemical isolation~\cite{Mocko15, Heinitz18, Dorrer18}.
In order to measure a high-statistics decay spectrum, as many of the produced $^{163}\mathrm{Ho}$ atoms as possible are required for the calorimetric measurement and only a minor fraction of that amount (max.~$10^{16}$ atoms corresponding to~$\sim\,2.7\,\mu\mathrm{g}$) is available for the direct $Q_{\mathrm{EC}}$-value measurement.

HCI can be produced and studied in electron beam ion traps (EBITs)~\cite{Levine88, Levine89} which allow the production of a large variety of different species and charge states up to even the highest charge states of heavy elements~\cite{Marrs94}.
EBITs are built for and operated at a broad range of electron beam energies starting from a few  hundred eV to a few hundred keV. 
The EBITs operated at higher electron beam energies typically employ a superconducting magnet which generates a magnetic field with a maximum strength of several Tesla for the compression of the electron beam.
In the last decades, also EBITs operated at room-temperature have been developed~\cite{Khodja97, Ovsyvannikov00, Micke18}. The magnetic field in these EBITs is created by means of permanent magnets and therefore less maintenance is required. The disadvantage is that the achievable vacuum pressures are higher than in cryogenic, superconducting EBITs resulting in a larger charge exchange rate and lower charge states.
For stable isotopes which are available in gaseous form or as volatile, organic compounds the injection into an EBIT is typically achieved using a differentially pumped injection system or by introduction of a tiny leak into the vacuum system through which the species of interest are injected into the background gas.
However, since the background gas in the vacuum system is flooded with the injected gas, a large quantity of the species introduced in the background gas is again pumped out of the system without being ionized and trapped by the electron beam.
When HCI of very rare and radioactive species are required, a more efficient injection method has to be used in order to reduce the loss of sample material.
The \textsc{Pentatrap} experiment, requires, among others, HCI of the long-lived, synthetic radioisotope $^{163}\mathrm{Ho}$. Therefore a dedicated Heidelberg compact EBIT~\cite{Micke18} was constructed and an in-trap laser desorption source developed.
This paper reports on the development of an in-trap laser desorption technique which allows the production of HCI from very rare isotopes available in sample sizes down to $10^{12}$ atoms.


\section{Methods}
\label{sec:methods}

For the work presented in this paper a Heidelberg compact electron beam ion trap~\cite{Micke18} (HC-EBIT) was built for HCI production.
The inhomogeneous magnetic field needed for electron-beam compression is generated by 24~stacks of 3~permanent magnets each, and guided and focused by soft-iron elements resulting in a magnetic field of around 850~mT in the 2~cm long trapping region. 
With a maximum of 10~keV electron beam energy and 80~mA electron beam current a broad range of HCI in the medium to heavy mass region as well as bare nuclei of lighter elements are accessible.
The background vacuum pressure in this room-temperature EBIT is typically in the lower $10^{-9}$~mbar region.

Using laser ablation, it is possible to ablate small amounts of atoms and singly charged ions from a surface with about $10^{16}$ atoms of $^{163}\mathrm{Ho}$~\cite{Eibach14, Eliseev15}.
To utilize the possibility of using very small samples, the laser ablation technique is implemented in the HC-EBIT to produce HCI of $^{163}\mathrm{Ho}$.
In order to maximize the efficiency, the desorption process takes place at sub-mm distances from the electron beam in the trapping region using laser pulse energies below the ablation threshold.
This results in a much higher yield than the use of an external, dedicated laser ablation ion source and the subsequent transfer and capture of singly charged ions in the EBIT~\cite{Mironov03}, since laser ablation in those sources removes mostly neutral atoms from the surface and the small fraction of ions which is produced experiences losses in the aforementioned transfer and capture process in the EBIT.
Our method brings a substantial fraction of the ablated atoms to the electron beam while needing less laser pulse energy than the laser ion sources thereby reducing the amount of material ablated per laser shot.
Due to the lower power density on the target surface it is likely that the process in which atoms are removed from the surface is no longer an ablation process and will be named desorption process in the following.

\begin{figure*}
	\centering
		\includegraphics[width=17cm]{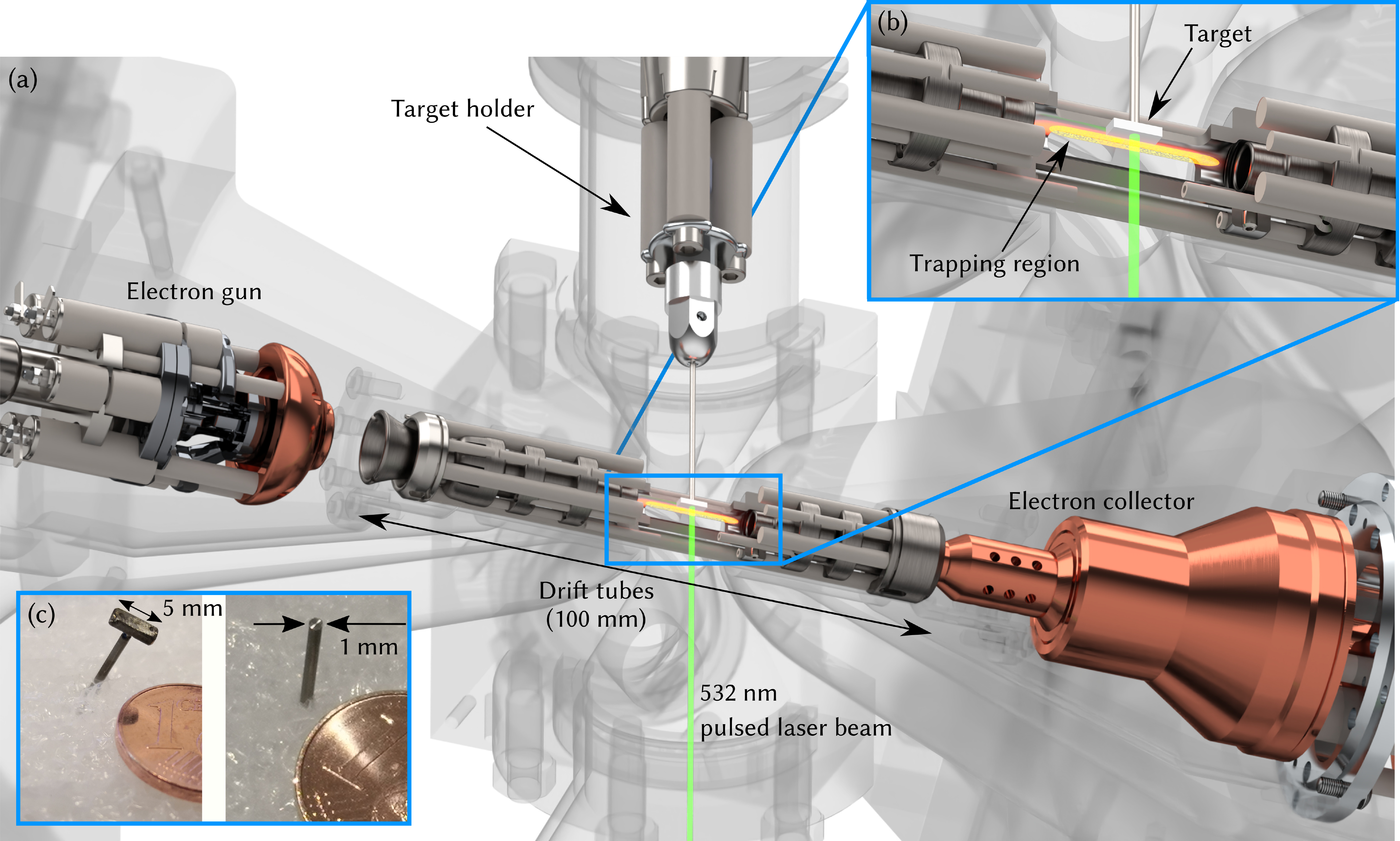}
	\caption{(a) Model of the HC-EBIT with the in-trap laser desorption setup. 
	The most relevant parts of the EBIT are the electron gun on the left, the stack of drift tubes in the center and the electron collector on the right side.
	In order to show the arrangement for in-trap laser desorption, the central drift tube is cut open. 
	Through a slit in the central drift tube the electrically insulated target holder is moved from above into the central drift tube near to the trapping region (position of the ion cloud shown in orange) using a three-axis, step-motor controlled manipulator. 
	The 532~nm pulsed laser beam (green) propagates along the vertical axis through the trapping region onto the target.
	This particular arrangement of trapping region, laser target, and laser beam is shown in the inset (b).
	(c) Pictures of the laser targets used for the commissioning of the in-trap laser desorption technique. 
	The left one shows the ``massive'' target with a thin foil of stable holmium spot-welded onto the target holder surface (2 x 5~mm). 
	The right side shows a laser target with $10^{12}$ atoms of $^{165}\mathrm{Ho}$ prepared on the surface of a 1~mm thick titanium wire using the ``Drop-on-Demand'' ink-jet printing technique~\cite{Haas17} as it was used for the presented measurements. 
	The dried drop of the diluted solution is not visible by eye.
	A 1~cent coin is shown for size comparison.}
	\label{fig:HC-EBIT}
\end{figure*}

A rendered model of the basic setup is shown in Figure~\ref{fig:HC-EBIT}.
The outer, transparent parts are the vacuum chamber and surrounding it the permanent magnets and soft-iron elements guiding the magnetic field and shaping it around the trapping region.
Inside the vacuum chamber, the electron gun is shown on the left, the electron collector on the right side and the stack of drift tubes between them. 
HCIs stored in the EBIT are ejected by pulsing down the trapping potential applied to the drift tube on the collector side.
Optical access to the trapping region is provided by four slotted apertures in the central drift tube (cut open in Figure~\ref{fig:HC-EBIT}).
From above, the target holder with the sample on the surface is lowered into the central drift tube very near to the trapping region (orange).
Its position is adjusted using a three-axis, step-motor controlled manipulator.
For desorption we use a pulsed, frequency-doubled Nd:YAG laser with a few mJ pulse energy and 7\,ns pulse duration at 532~nm pointed through a vacuum viewport at the target using a piezoelectric-driven mirror outside the vacuum.
The spot diameter on the target is approximately 300~to 400~$\mu$m and was estimated by inspecting the target surface with a scanning electron microscope after it was used.
Attached to the two remaining access ports to the trapping region (not visible in Figure~\ref{fig:HC-EBIT}) are an x-ray detector (Ketek AXAS-D Vitus H50) and a gas injection setup.

\begin{figure}
	\centering
		\includegraphics[width = 8.5cm]{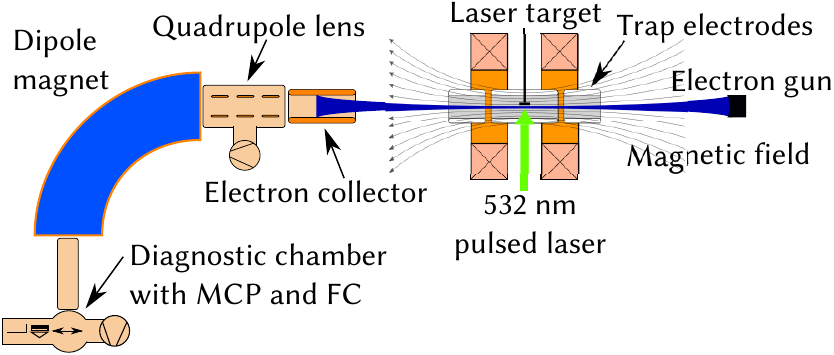}
	\caption{Overview of the test setup for the commissioning of the in-trap laser desorption technique. 
	The right side of the figure schematically shows the main parts of the HC-EBIT as described in Figure~\ref{fig:HC-EBIT} including the laser beam (green arrow) and the target (black).
	Ions ejected from the HC-EBIT pass through the electron collector and a quadrupole lens which is used for beam steering and focusing. 
	The dipole magnet separates the ion bunches according to their charge-to-mass ratio and the selected ion species is subsequently detected in the detector chamber fitted with a microchannel plate (MCP) detector and a Faraday cup (FC).}
	\label{fig:TestSetup}
\end{figure}

For the identification of the produced ions and charge states a diagnostic setup consisting of a quadrupole lens, a dipole magnet and two detectors (microchannel plate detector (MCP) in Chevron configuration with a phosphor screen and a camera as well as a Faraday cup (FC)) is attached to the HC-EBIT as shown in Figure~\ref{fig:TestSetup}.
By detecting the ion signal on the MCP as a function of the field of the dipole magnet while scanning this parameter, an isotopically resolved charge-state spectrum of the produced species is obtained.
The gain setting of the MCP was reduced in order to only detect HCI and suppress background from ions in lower charge states.
Initially, a reference species has to be measured in order to calibrate the dipole magnet, where in our case xenon gas (natural isotopic composition) is injected into the background gas of the HC-EBIT using a gas dosing valve.
In the obtained spectrum of natural xenon the overlap of xenon isotopes from adjacent charge states allows the identification of the charge states and calibration of the magnetic field to a charge-to-mass ratio.

Different targets (cf. Fig. \ref{fig:HC-EBIT}~(c)) containing the stable isotope $^{165}\mathrm{Ho}$ were used for commissioning (natural holmium is monoisotopic). 
The target holder is manufactured from titanium and consists of a wire (1 mm diameter) and a 2x5~$\mathrm{mm}^2$ titanium plate welded to its end. 
Initial tests were performed with a thin holmium foil which was spot-welded to the target holder giving a sample size that is almost infinite for our purposes. 
These tests showed that the operation of the HC-EBIT remains stable while a laser pulse is fired onto the target very close to the trapping region in the EBIT and that desorbed atoms or ions are subsequently trapped and ionized to high charge states and remain trapped.
The produced HCI of stable holmium were extracted and the charge states identified using the test setup (cf. Figure \ref{fig:TestSetup}) as described above.

In the following, we used targets with a specific total number of holmium atoms deposited on the surface.
For sample preparation the ``Drop-on-Demand'' ink-jet printing technique~\cite{Haas17} is used where a diluted standard solution of stable holmium is printed into one spot of the target.
Note that the printed drop has a volume of just 8~nL and is visually not recognizable on the target surface (cf. Fig. \ref{fig:HC-EBIT}~(c)).
A total uncertainty of 3\,\% of the number of atoms in the sample results from uncertainties of 2\,\% in the drop volume, 2\,\% in the solution concentration and 1\,\% due to aging of the solution, respectively.
Since a smaller target surface significantly facilitates the adjustment of the laser spot onto the invisible sample, only the 1~mm diameter titanium wire (flattened on one side using a lathe) was used for the smallest samples.
Titanium was chosen as target holder material since it is a comparably light element and titanium HCI reach only lower charge states than the much heavier holmium HCI. This leads to selective evaporative cooling of titanium HCI and favors the accumulation of holmium HCI~\cite{Schneider89}.

For an efficient injection of atoms from the sample into the EBIT's trapping volume, the position of the target and the applied bias voltage are crucial.
Fine positioning of the target into its final position is performed with the HC-EBIT in operation in order to monitor any changes in its performance caused by the target as it approaches the trapping region and the electron beam.
At about 20~mm from the trapping region, the potential of the target is set to the same potential as the central drift tube to avoid a deflection of the electron beam and discharges when moving through the slot in the central drift tube.
From here on the target position in relation to the drift tubes is observed with a camera into the propagation direction of the laser.
With the target inside the central drift tube, steps of about 100~$\mu$m are used to move it towards the electron beam.
Following each step, the count rate on the x-ray detector and its spectrum are monitored to avoid moving the target into the electron beam which causes a considerable amount of bremsstrahlung.
As soon as the count rate increases, the target is moved back by 100~$\mu$m and the bias voltage applied to the target is adjusted for the maximum extracted ion signal on the MCP.
This ensures that the HCI-trapping potential is not perturbed by the presence of the target holder while reaching a minimum distance between the laser target and the electron beam.

For the extraction of ion bunches, the HC-EBIT is operated in cycles where the inventory of trapped HCI is extracted at regular intervals.
In each cycle, the loading with a single laser pulse and HCI breeding are followed by a pulsed lowering of the potential of the trap electrode closest to the collector.
The time interval when the trap is closed is referred to as the charge breeding time $t_{\mathrm{br}}$ and the short time during which the drift tube is pulsed to a lower potential as the ejection time $t_{\mathrm{ejec}}$.
For the measurements presented in the next section, the HC-EBIT was operated with a cycle time of 1.001~s of which the last $t_{\mathrm{ejec}}= 5\,\mu\mathrm{s}$ were reserved for the ion extraction.
The laser pulse was triggered 1 ms after the cycle starts, leaving about $t_{\mathrm{br}} = 1\,\mathrm{s}$ for charge breeding of the injected atomic species.

For a new target, the lowest possible laser pulse energy setting should be used for positioning the laser spot onto the region where the species of interest is presumed to be on the target. 
This pulse energy setting can be found by monitoring the x-ray spectrum while the laser pulse energy is increased until a characteristic x-ray line of the injected species (shifted to higher energy since the species is quickly ionized to high charge states) is observed.
The laser positioning procedure is monitored again using the camera and thereby ensuring that the laser spot stays on the target surface while adjusting its position.
Once the laser spot is positioned, the dipole magnet is tuned to a current setting which should guide HCI of the injected species to the MCP.
Typically, the first atoms are removed from the target surface and injected into the trap when the laser pulse energy reaches 1~mJ.
Within a few trap cycles after the laser is switched off, the injected species completely vanishes from both, the x-ray spectrum and the MCP resulting in an accurately controllable injection leaving no contamination in the background gas.


\section{Results}
\label{sec:results}

The setup presented in Figure \ref{fig:TestSetup} was used to test the reliability of the HC-EBIT with in-trap laser desorption and to identify the HCI species in the extracted ion bunches by recording isotopically resolved charge-state spectra when scanning the magnetic field of the dipole magnet.
For each current setting only HCI with a specific charge-to-mass ratio can pass the dipole magnet and are detected by one of the detectors in the diagnostic chamber located after the magnet.
For each scan a few thousand extracted ion bunches were needed to obtain a full spectrum of the produced charge states.

\begin{figure}
	\centering
		\includegraphics[width = 8.5cm]{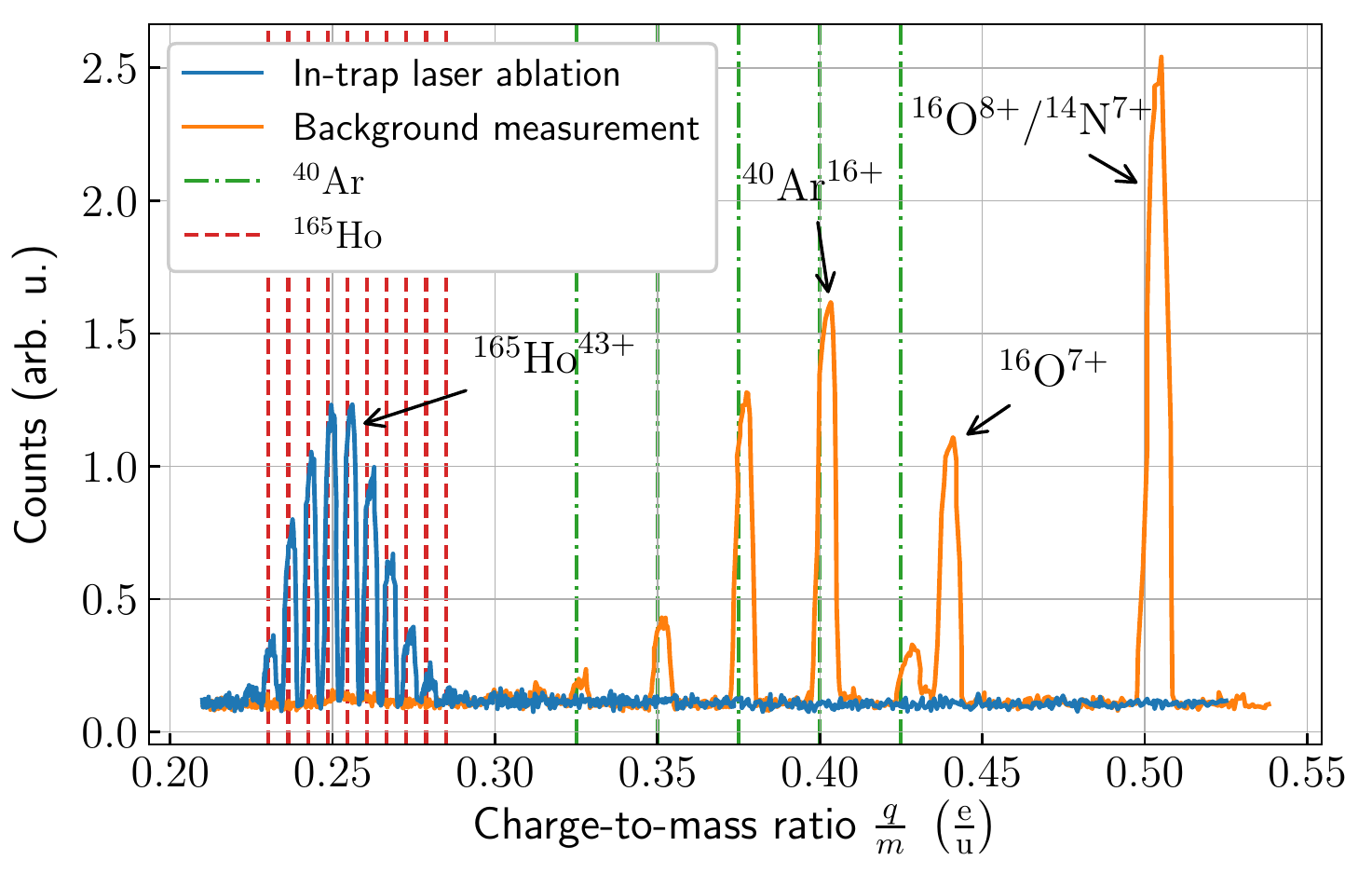}
	\caption{Measured distribution of holmium charge states (blue) using a foil of $^{165}\mathrm{Ho}$ as laser target and background measurement (orange) without laser desorption. 
	Dashed lines indicate the expected positions of holmium (red) and dash-dotted lines argon charge states (green). 
	The blue spectrum exclusively shows peaks which coincide with the expected positions of holmium charge states centered around $^{165}\mathrm{Ho}^{42+/43+}$. 
	Otherwise no background was observed and holmium seems to be the sole species in the trap. 
	In the measurement without laser desorption we observed mainly argon in high charge states as well as light, fully ionized rest gas atoms such as oxygen and nitrogen at a charge-to-mass ratios around~0.5.}
	\label{fig:MassiveTarget}
\end{figure}

For the first tests a massive target made of solid natural holmium foil (mainly $^{165}\mathrm{Ho}$) was used.
A measured spectrum of $^{165}\mathrm{Ho}$ charge states obtained with this target is shown in Figure~\ref{fig:MassiveTarget}.
The blue curve was measured with laser desorption while the orange curve shows a background measurement without laser desorption but with the target still positioned in the trapping region. 
As long as holmium was injected into the EBIT the spectrum was essentially free of any background since the relatively heavy holmium ions experience a deeper trapping potential than the lighter, lower charged ions from the residual gas.
These evaporate more easily from the trap and thereby cool the holmium HCI.
This measurement demonstrates the successful implementation of the in-trap laser desorption technique and also that the desorption process does not perturb the HCI production or their trapping.

\begin{figure}
	\centering
		\includegraphics[width = 8.5cm]{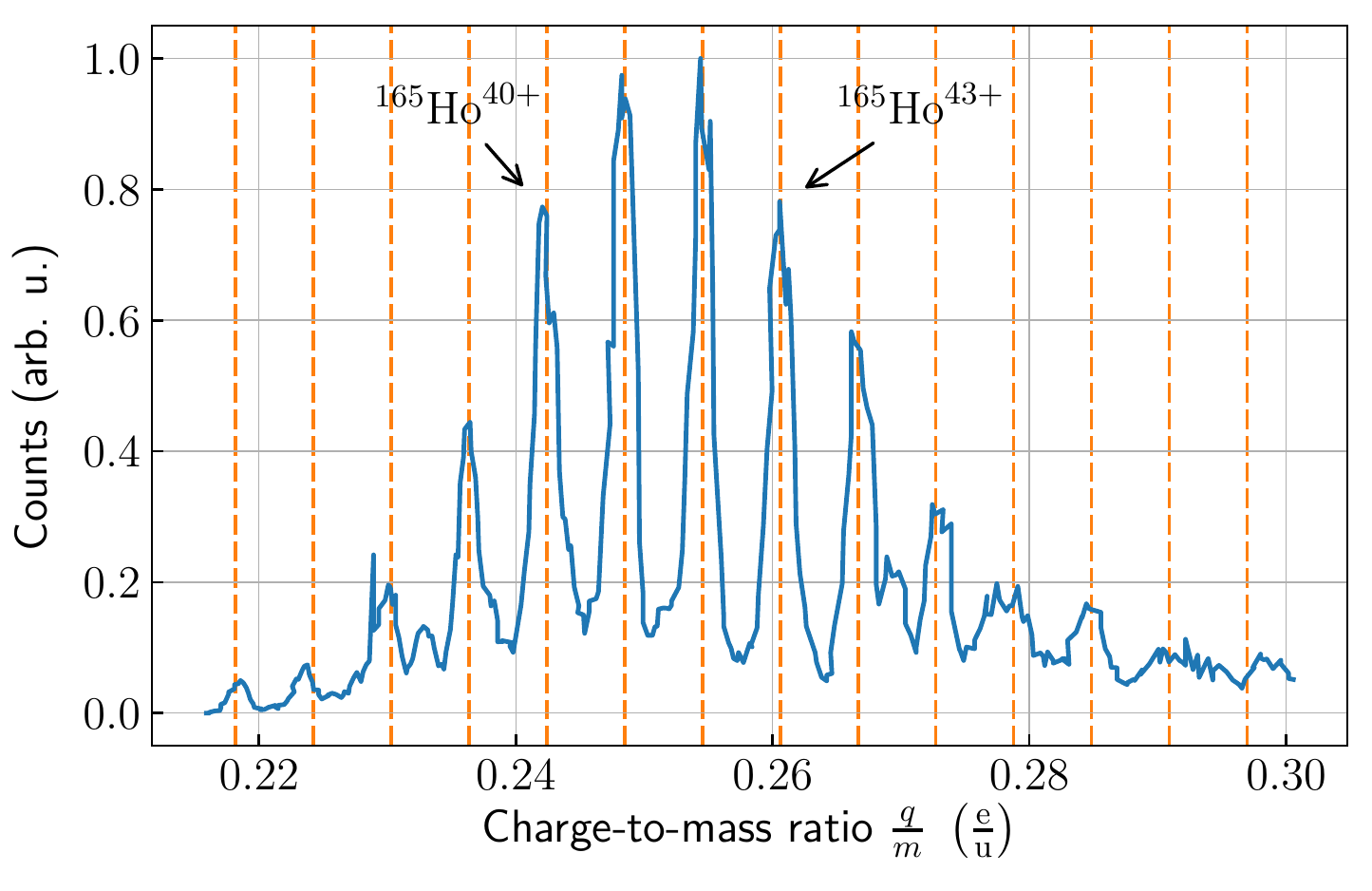}
	\caption{Charge state distribution of holmium HCI extracted from the HC-EBIT following injection by in-trap laser desorption from a target with $10^{12}$ atoms. 
	About 8 charge states centered around $^{165}\mathrm{Ho}^{42+}$ were observed.
	The dashed lines mark the charge-to-mass ratios where the holmium HCI were expected to appear after calibration of the dipole magnet.}
	\label{fig:ChargeStateDist}
\end{figure}

In order to find the practical limit in sample size of this injection technique, samples with a decreasing amount of stable holmium ($^{165}\mathrm{Ho}$) were tested.
For each new sample a spectrum of charge states was measured to verify the presence of holmium and exclude possible contaminants which are easily possible for these sample sizes.
The smallest sample size that could be reliably used so far contained $10^{12}$ stable holmium atoms, corresponding to a quantity of 300~pg. 
For each spectrum we needed about 3500 to 5000 laser shots, mainly depending on the time needed to optimize ion ejection (about 500 to 1500 laser shots) and the current range over which the dipole magnet was scanned.
A typical result for this sample size is shown in Figure~\ref{fig:ChargeStateDist}.
The spectrum was measured at a charge breeding time of $t_{\mathrm{br}} = 1\,\mathrm{s}$, an electron beam energy of 5.9~keV and 45~mA electron beam current with laser pulses of up to 4~mJ.
In comparison to a continuously injected species, only a narrow distribution of charge states was observed since the pulsed injection once per trap cycle lets all ions experience the same charge breeding time and therefore no ions in lower charge states were present.
Moreover, with $t_{\mathrm{br}} = 1\,\mathrm{s}$ the presented charge state distribution had already reached its equilibrium and a longer $t_{\mathrm{br}}$ would not have changed the distribution.
With an improved pumping system, recombination by charge-exchange would be better suppressed narrowing the charge state distribution even further.

With one sample it was possible to measure the charge-state distribution several times without observing a reduction in the intensity of the ion spot on the MCP.
In order to characterize the ``durability'' of the targets a measurement of the target lifetime was performed using again the smallest reliably tested sample size of $10^{12}$ stable holmium atoms, although conceivably lower limits could be achieved if needed for $Q_{\mathrm{EC}}$-measurements. 

\begin{figure}
\centering
	\includegraphics[width = 8.5cm]{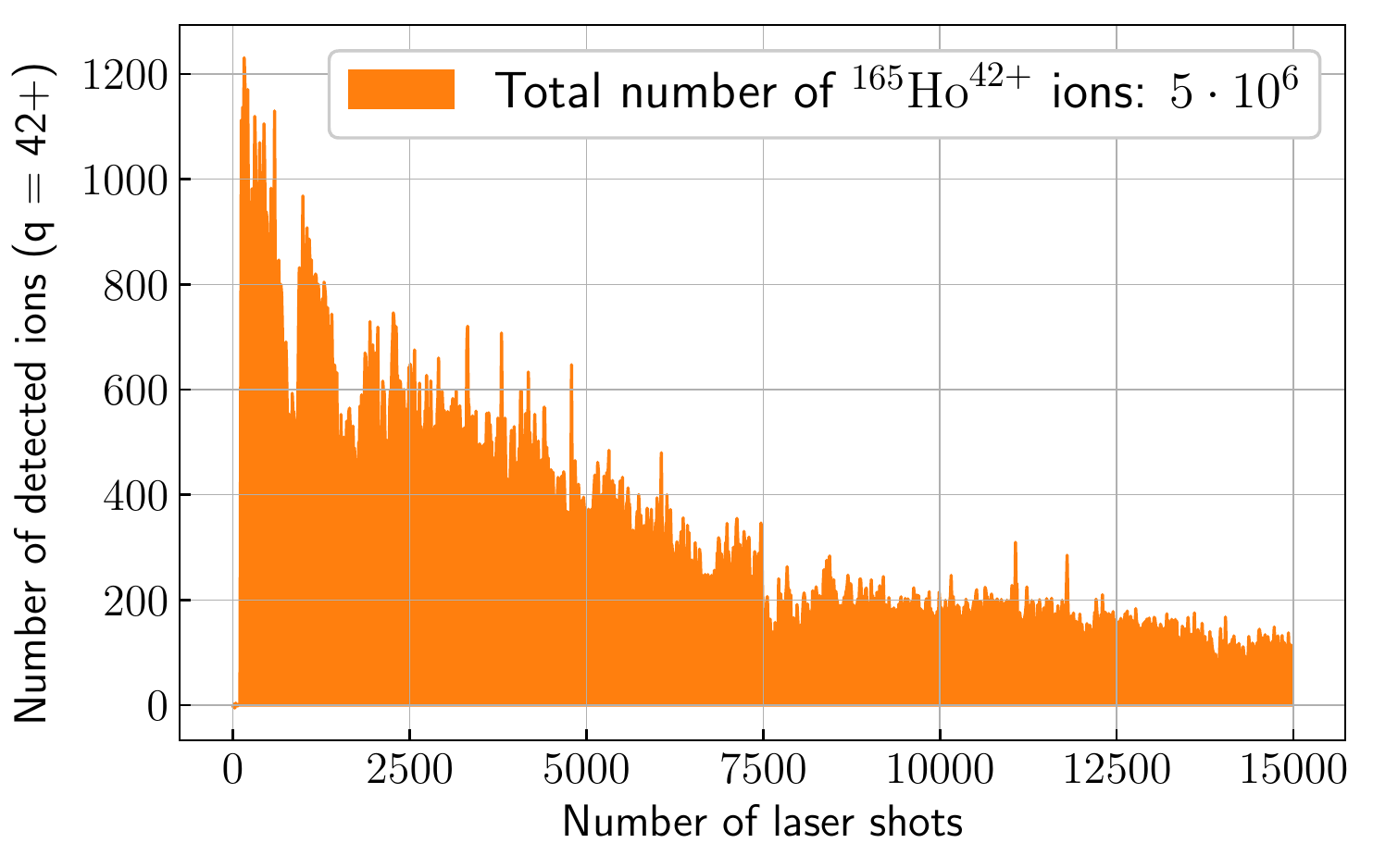}
\caption{Measurement of the lifetime of a sample with $10^{12}$ atoms of $^{165}\mathrm{Ho}$ using an MCP detecting only HCI in the charge state 42+ following a charge-to-mass separation with the dipole magnet. The total number of extracted HCI was about $5\cdot 10^{6}$ with 10~\% uncertainty.}
\label{fig:TargetLifetime}
\end{figure}

Experimentally, we measured the target lifetime by tuning the dipole magnet to the current setting for the most abundant charge state $^{165}\mathrm{Ho}^{42+}$ (cf. Figure~\ref{fig:ChargeStateDist}). 
Before the lifetime measurement was started the number of ions per laser shot was measured using the FC and a charge amplifier (Femto HQA-15M10T, Gain: 10~V/pC).
Then we continued with firing laser pulses onto the target until the holmium HCI in this charge state were not visible anymore, indicating the depletion of the target at the laser spot location).
Finally the number of ions was again measured using the FC and the lifetime measurement with the MCP was cross-calibrated against the two FC measurements at the beginning and at the end of the lifetime measurement.
An exemplary dataset of such a measurement is shown in Figure \ref{fig:TargetLifetime} which was acquired following two measurements of the charge-state distribution. It was obtained using 4~mJ laser pulses focused on a single spot on the target surface. 
Increasing the pulse energy as well as moving the laser spot can increase the ion yield again if the laser spot is smaller than the sample area.
Including the two measurements of charge state spectra such as the one shown in Figure \ref{fig:ChargeStateDist} (8000 laser pulses) and the lifetime measurement (15000 laser pulses) for a total of 23000 laser shots HCI of holmium in the charge state 42+ could be extracted from a target with $10^{12}$ atoms of $^{165}\mathrm{Ho}$.
By integration of the curve in Figure \ref{fig:TargetLifetime} the total number of extracted HCI in this charge state was estimated to have been around $5\cdot10^{6}$ with about 10~\% uncertainty resulting from the analysis of the MCP data.
This number takes only the ion number during the lifetime measurement into account and not the extracted ions during the measurements of the charge state spectra.

The given numbers depend strongly on the EBIT settings, electron beam current, target position, laser spot position on the target and the beamline settings and are therefore not exactly reproducible, i.e. the laser spot position might have to be reoptimized during ion extraction. 
Also the exact location of the sample on the target is not visible, therefore it was not possible to verify that the laser spot exactly coincided with the sample material on the target. 
Hence, the given numbers are estimates for only one laser spot position and can vary for different settings and positions.
When the laser spot position is changed the number of ions that are detected can increase again.
However, extracted ions are always observed once the laser spot is on the sample and the spot position can be subsequently optimized.


\section{Discussion and Conclusion}
\label{sec:discussion}

The presented experiments demonstrate injection, charge breeding and extraction of rare species HCI with a HC-EBIT by means of in-trap laser desorption using extremely small samples containing on the order of $10^{12}$ atoms.
Compared to the previously used dedicated laser ion source producing singly charged ions which are then injected into the EBIT for charge breeding \cite{Mironov03}, our method improves the efficiency by several orders of magnitude since desorbed neutral atoms are also captured in the EBIT and ion-transport losses between laser-ion source and EBIT are eliminated altogether.
The results were reliably obtained with holmium samples of $10^{12}$ atoms which lasted for a reasonable number of laser shots as required for high-precision mass-ratio measurements at the \textsc{Pentatrap} experiment.
Our technique can be further used to produce HCI of any other species, especially also of sufficiently long-lived, medium heavy synthetic radionuclides and transuranium elements\cite{Eibach14} produced in nuclear reactors and only available in small quantities.
The thereby produced HCI can be extracted as in our case but it is also possible to perform spectroscopic measurements on the HCI in the EBIT using either electron impact excitation or external radiation sources, e.g. synchrotrons or x-ray lasers for electronic excitation~\cite{Micke18}.

In comparison with the wire probe method\cite{Elliott95}, the use of a laser pulse triggering the trap loading allows a much better experimental control and is better adapted to experiments requiring regular ion extraction.
To compare the performance of the two methods one has to consider that the spectroscopic measurements on uranium isotopes with the wire probe method did not require the ejection of HCI but ions accumulated for several minutes and stored for several hours which reduces the consumption of sample material.
The smallest used sample of uranium consisted of about $10^{14}$ atoms of $^{233}\mathrm{U}$.
For spectroscopic measurements of this type just one single laser pulse is required to load the trap and the ions can then be stored in the EBIT for a similarly long measurement time using samples with two orders of magnitude fewer atoms.
In principle, if the intended experiments can be tuned and optimized using a more abundant HCI of the same charge-to-mass ratio, then even smaller sample sizes towards the $10^{10}$ atoms region are conceivable.

With the present technique, the range of rare isotopes that can be made available for high-precision nuclear mass measurements and spectroscopic experiments has been substantially expanded, and the use of samples of picogramme size has become a real possibility with interesting consequences for nuclear as well as atomic physics and other fundamental studies.


\begin{acknowledgements}
This project has received funding from the \textit{European Research Council (ERC)} under the European Union's Horizon 2020 research and innovation programme under grant agreement No. 832848 - FunI.
Furthermore we acknowledge funding and support by the Max-Planck Society, the international Max-Planck research school for precision tests of fundamental symmetries (IMPRS-PTFS) and funding from the DFG Research UNIT FOR 2202 under project number DU1334/1-2.
\end{acknowledgements}


%
%

%

\bibliography{InTrapLaserDesorption}

\end{document}